\documentclass[11pt,twoside]{article}
\usepackage{asp2010}

\resetcounters

\begin{document}

\title{Planet Formation Around M-dwarf Stars \\ From Young Disks to Planets}
\author{Pascucci, I.$^1$, Laughlin, G.$^2$,  Gaudi, B. S.$^3$, Kennedy, G.$^4$, Luhman, K.$^5$, Mohanty, S.$^6$, Birkby, J.$^4$, Ercolano, B.$^4$, Plavchan, P.$^7$, Skemer, A.$^8$ 
\affil{$^1$Space Telescope Science Institute, Baltimore, USA}
\affil{$^2$University of California, Santa Cruz, USA}
\affil{$^3$The Ohio State University, USA}
\affil{$^4$Institute of Astronomy, Cambridge, UK}
\affil{$^5$The Pennsylvania State University, USA}
\affil{$^6$Imperial College London, UK}
\affil{$^7$NASA Exoplanet Science Institute, USA}
\affil{$^8$University of Arizona, Steward Observatory, USA}
}

\begin{abstract}
Cool M dwarfs outnumber sun-like G stars by ten to one in the solar
neighborhood. Due to their proximity, small size, and low mass, M-dwarf stars are
becoming attractive targets for exoplanet searches via almost all current search
methods. But what planetary systems can form around M dwarfs?
Following up on the Cool Stars~16 Splinter Session "Planet Formation Around M Dwarfs", we summarize here our knowledge 
of protoplanetary disks around cool stars, how they disperse, what planetary systems 
might form and can be detected with current and future instruments.  
\end{abstract}

\section{Disk Observations}

\subsection{The Properties of Disks Around M Dwarfs}\label{sect:properties_disks}
There is growing observational evidence that young (1-2 Myr) very low-mass stars (M
dwarfs) and brown dwarfs undergo a T Tauri phase: they are surrounded by optically thick
gas-rich dust disks, they accrete disk gas, and have jets/outflows like their more
massive counterparts. {\bf Subhanjoy Mohanty} from the Imperial College, London,
presented an extensive overview of the properties of these young very low-mass stars and
brown dwarfs. He reviewed the disk accretion and the jet diagnostics detected around
young cool stars with special emphasis on the broad ($\sim$100\,km/s) H$\alpha$
emission lines (e.g., \citealt{2005ApJ...626..498M}), and the broad and blueshifted
forbidden [OI] line at 6300\,\AA{} (e.g., \citealt{2009ApJ...706.1054W}). He pointed out that accretion and infall are detected even in $\sim$10\,Myr-old very low-mass stars and brown dwarfs, similarly to what is found for sun-like stars \citep{2003ApJ...593L.109M,2009ApJ...696.1589H,2010ApJ...714...45L}. There also appears to be a strong correlation between the mass of the central star and the amount of gas accreted onto it ($\dot{M} \propto M_{\star}^2$, \citealt{2003ApJ...592..266M,2006A&A...452..245N,2008ApJ...681..594H}), a correlation which challenges our theoretical understanding of viscous disk evolution. Infrared emission in excess to the stellar photosphere is  often detected around young very low-mass stars and brown dwarfs pointing to the presence of circumstellar disks within which planets might form. The disk frequency of $\sim$50\% is found to be similar to that around young sun-like stars indicating that the raw material for planet formation  is often available regardless of the star mass (e.g., \citealt{2007prpl.conf..443L}).

In spite of similar diagnostics for the T Tauri phase, differences are emerging in the properties of disks as a function of stellar mass. Dust disks around very low-mass stars and brown dwarfs appear to have smaller scale heights than disks around sun-like stars of similar age \citep{2010ApJ...720.1668S}. This could result from a real difference in the disk structure, due for instance to faster grain growth and hence decoupling of the dust from the gas followed by settling, or from an opacity effect caused by infrared observations probing closer to the disk midplane in disks around cool stars \citep{2007ApJ...659..680K}. Supporting the first scenario, dust grains in the disk atmosphere traced via prominent silicate emission features at 10 and 20\,\micron{} appear to be more evolved around brown dwarfs than around sun-like stars (\citealt{2005Sci...310..834A,2006ApJ...639..275K,2009ApJ...696..143P} and Sect.~\ref{sect:evolution}).

The outer disk radius and disk mass are among the most important parameters to assess
the likelihood of forming planets around very low-mass stars and brown dwarfs (see
Sect.~\ref{sect:pftheory}). Mohanty presented preliminary results from a large
JCMT/SCUBA-2 survey which, combined with previous measurements, suggest that: i) M$_{\rm
disk}$/M$_{\rm star}$ at a given age (e.g., Taurus) is consistent with being constant
from brown dwarfs to sun-like stars, at $\sim$1\%, but ii) the same ratio appears to
significantly decrease with increasing age (e.g., by TWA).  The latter finding is
consistent with a substantial increase in grain size/settling with age in brown dwarf
disks, as noted previously for sun-like stars. Disk radii are notoriously
difficult to determine. \citet{2009A&A...496..725E} recently proposed to use the ratio
of two forbidden [CI] lines, one in the near- and the other in the far-infrared, as a diagnostic for outer disk radii. The first near-infrared [CI] line detection is now available toward TWA30B \citep{2010ApJ...714...45L}. Far-infrared observations of [CI] lines at 370 and 609\micron{}  will be possible with the Herschel Space Observatory.

One system of special interest is the brown dwarf 2MASS1207 in the TW~Hya association, which is surrounded by a dust disk and has a planetary mass companion (2MASS1207b) at a projected separation of  $\sim$50\,AU. 2MASS1207b is a peculiar object in that it is under-luminous by two orders of magnitudes than expected at all bands from I to L. {\bf Andy Skemer} from the University of Arizona summarized what scenarios have been proposed to explain its under-luminosity and evaluated their likelihood. Based on modeling of the full spectral energy distribution (SED), and absence of photometric variability, he argued that gray extinction by a nearly edge-on disk is highly unlikely. He pointed to clouds of micron-sized dust grains in the atmosphere of 2MASS1207b as a possible sink of luminosity. Unfortunately, current atmospheric models do not treat self-consistently the effect of dust grains in planets' atmospheres, hence a quantitative comparison between theory and observations is yet not feasible.

\subsection{Disk Evolution around Sun-like Stars and M Dwarfs}\label{sect:evolution}
As summarized in the previous section, very low-mass stars and brown dwarfs undergo a T Tauri phase that in some aspects resembles that around sun-like stars. In this part of the CS16 session, {\bf Kevin Luhman} from The Pennsylvania State University addressed the question of how disks evolve and disperse with emphasis on similarities and differences between sun-like stars and M dwarfs/brown dwarfs. He showed that the  frequency of infrared excess emission appears to decrease less steeply with age as the mass of the central object decreases, although this result   is not yet conclusive because few disk fractions have been measured for low-mass stars older than 3 Myr \citep{2006ApJ...651L..49C,2009AIPC.1094...55L}. If this trend is confirmed the implication is that optically thick dust disks, providing the raw material to form planets, persist longer around  very low-mass stars and brown dwarfs than around sun-like stars. 

\begin{figure}[!ht]
\plotone{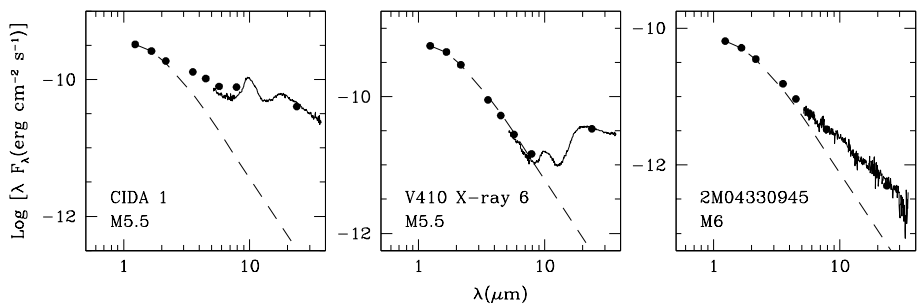}\label{fig:sampleSEDs}
\caption{Sample SEDs around very low-mass stars and brown dwarfs. Left panel: SED of a primordial optically thick dust disk. Middle panel: SED of a transitional disk. Right panel: SED of an evolved/settled disk.}
\end{figure}

Detailed studies of SED shapes demonstrate that very low-mass stars and brown dwarfs undergo the same disk clearing phases as their higher mass counterparts. There is evidence for transitional disks (reduced mid- but large far-infrared emission caused by a gap or hole) around brown dwarfs as well as for settled disks (reduced emission at all infrared wavelengths), see Fig.~\ref{fig:sampleSEDs} and \cite{2006ApJ...643.1003M}. The fraction of disks with transitional SEDs is about $\sim$15\% in disks around sun-likes stars as well as M dwarfs in young (a few Myr-old) star-forming regions. The paucity of transition disks in both star samples point to a rapid disk clearing, on a timescale of $\sim$0.1\,Myr \citep{2010ApJS..186..111L}.

In addition to the dispersal of primordial material, brown dwarf disks also present the same signs of dust processing as disks around sun-like stars. Grain growth as well as crystalline grains have been detected in brown dwarf disks \citep{2005Sci...310..834A,2005ApJ...621L.129F} with some evidence for more efficient processing than in disks around sun-like stars belonging to the same star-forming region \citep{2007ApJ...660.1517S,2008ApJ...676L.143M,2009ApJ...705.1173R}. There is also some hint that grain growth may be affected by the presence of sub-stellar companions (Adame et al. submitted).

Finally, thanks to the sensitivity of new infrared and UV instrumentation, we are
starting to have detections of gaseous species in disks around very low-mass stars and brown dwarfs \citep{2009ApJ...696..143P,2010ApJ...715..596F}. Differences in the column densities of the detected species might result from the different stellar radiation field impinging onto the disk surface \citep{2009ApJ...696..143P}. It will be interesting to determine how much the chemistry of the disk midplane is impacted by this different radiation field and what are the effects on the composition of forming planets.

\begin{figure}[!ht]
\plottwo{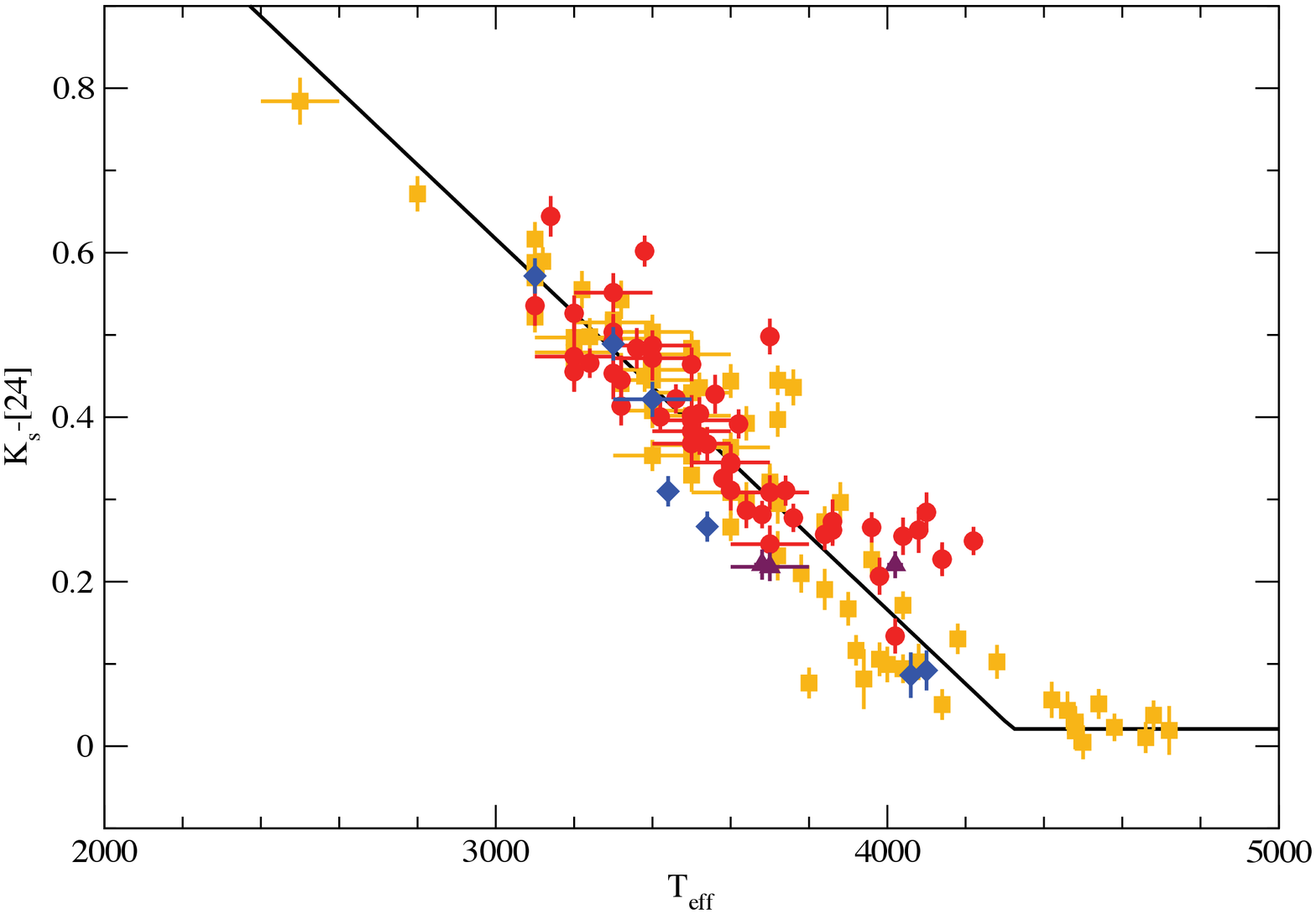}{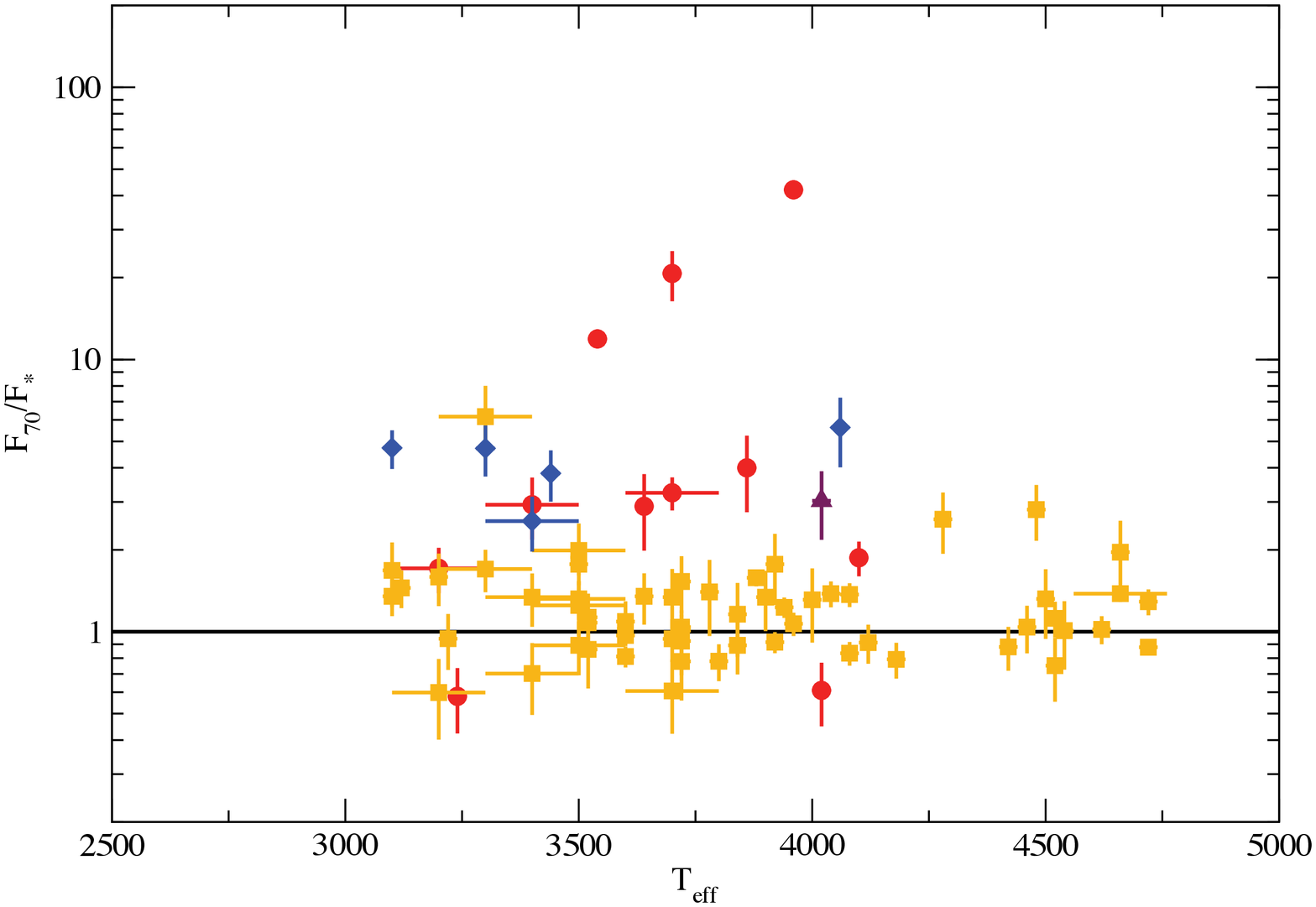}\label{fig:debris}
\caption{Spitzer 24 and 70 micron excess plots for the following samples: the control sample of nearby, older M dwarfs (yellow), Xray active M dwarfs (red), rapid rotators (blue) and IRAS and sub-mm excess candidates (purple).  These plots show that only Xray active M dwarfs tend to have a higher frequency of small warm 24\micron{} excesses, while both Xray active and rapid-rotator M dwarfs tend to have a higher disk frequency at 70\micron{} when compared to nearby, older M dwarfs.}
\end{figure}

Moving into older/nearby M dwarfs, {\bf Peter Plavchan} from the NASA Exoplanet Science Institute reported on his latest results from a search of debris dust disks, second generation dust disks, around M-dwarf stars. He showed that there is a decreasing rate of debris disks detections with spectral type going from A stars to FGK stars down to the M dwarfs \citep{2009ApJ...698.1068P}. Because at young ages ($<$200\,Myr) the frequency of debris disks around M dwarfs is similar to that around old sun-like stars, there could be an efficient mechanism to remove debris dust around M-dwarf stars.  \citet{2009ApJ...698.1068P} proposed stellar wind drag as the most likely mechanism. They are currently testing this scenario via a large survey of debris dust disks with Spitzer around  Xray bright M dwarfs likely to be young and not having large radial stellar winds. Preliminary results confirm that Xray bright M dwarfs have detectable debris disks with a frequency that is very similar to that of debris around sun-like stars (Plavchan et al. in prep. and Fig.~\ref{fig:debris}). Because debris dust is very likely linked to the formation of planetesimals, these results indicate that planetesimal formation occurs as often around M-dwarf stars as around sun-like stars.

\section{Theory}
After discussing the observational properties of young and old disks around M dwarfs, we turned to the theory of disk dispersal and planet formation with the goal of understanding what type of planetary systems can form around M-dwarf stars.

\subsection{Theory of Disk Dispersal around M Dwarfs}
\begin{figure}[!ht]
\epsscale{0.1}
\plotone{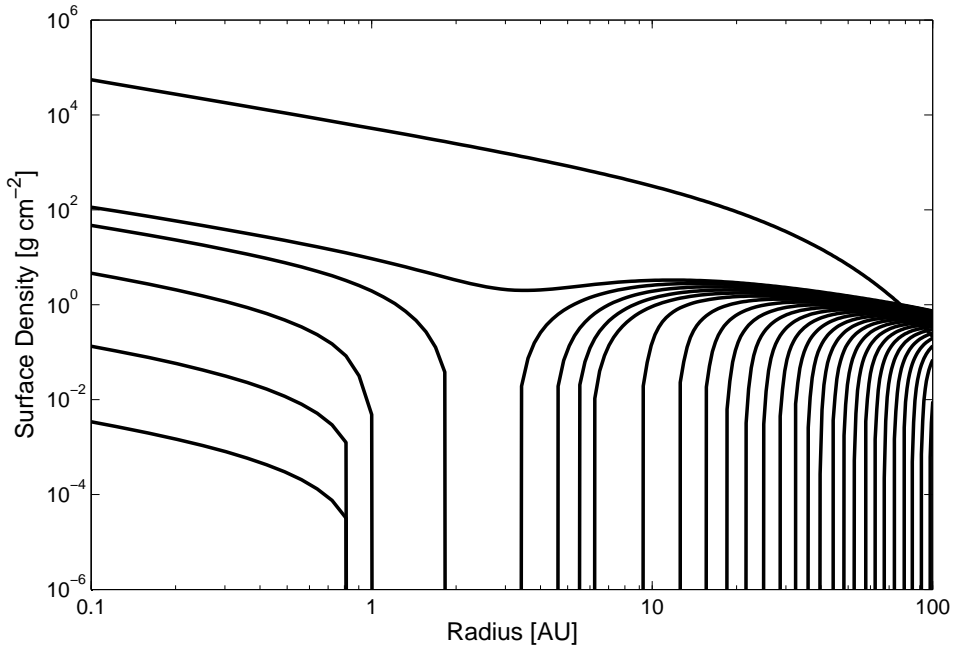}\label{fig:photevap}
\caption{An example of disk dispersal due to X-ray photoevaporation and viscous evolution. The
first (highest) line shows the initial surface density profile, the next shows the profile
at 75\% of the disk lifetime. Remaining lines show the surface density at 1\% steps in disk
lifetime. Figure courtesy of James Owen, see also \citet{2010MNRAS.401.1415O}.
}
\end{figure}

{\bf Grant Kennedy} from the Institute of Astronomy, Cambridge, reviewed in detail the physical mechanisms dispersing protoplanetary disks and how they might act to remove the primordial gas and dust around M dwarfs. 
Viscous accretion of gas onto the central star, accompanied by outward spreading, is certainly a major player in disk evolution. As summarized in Sect.~\ref{sect:properties_disks} young M dwarfs and even brown dwarfs experience disk mass accretion in the first few Myr of their lives. Although ubiquitous, viscous evolution alone cannot reproduce the relatively short disk lifetimes of 1-10\,Myr inferred from observations, as noted e.g. by \citet{2001MNRAS.328..485C}. Photoevaporation driven by the central star has been proposed as the next ubiquitous and most relevant disk dispersal mechanism.
In brief, radiation from the central star (X-ray, EUV and FUV photons) heat the gas in the upper layers of the disk, depositing enough energy to leave the gas unbound, thus establishing a photoevaporative flow (\citealt{2000prpl.conf..401H,2007prpl.conf..555D} for reviews). 
When the accretion rate drops below the wind loss rate, photoevaporation limits the supply of gas to the inner disk, which drains onto the star on the local viscous timescale--- of order 10$^5$ years (see Fig.~\ref{fig:photevap}). With the inner disk gone, the remaining outer disk is rapidly dispersed by direct stellar radiation. Though the original model of Clarke et al. invoked EUV photons as the main cause of photoevaporation (see also \citealt{2006MNRAS.369..216A} for following developments), recent work suggests that FUV and Xray radiation, which launch winds from denser disk regions, are more important \citep{2008ApJ...688..398E,2009ApJ...699.1639E,2009ApJ...705.1237G}.
%When the accretion rate drops below the wind loss rate, photoevaporation takes over and carves %a hole within the gas disk, thus decoupling the inner and outer disks. The inner disk drains %onto the star on a viscous timescale, just 100,000 years, while the outer disk is blown away %from photoevaporation on a timescale that sensitively depends on the dominant stellar %high-%energy photons (REF). 
The first strong observational evidence for photoevaporation driven by the central star has been recently presented by \citet{2009ApJ...702..724P} for the disk around the $\sim$10\,Myr-old star TW~Hya. 

\citet{2009ApJ...690.1539G} modeled the dispersal of protoplanetary disks around stars of different masses and different radiation fields. They found that the disk lifetime substantially decreases for stars more massive than 7\,M$_\odot$, due to their high UV fields, but does not vary much for stellar masses in the range 0.3-3\,M$_\odot$, hence from M dwarfs up to a few times the mass of the Sun. This is likely because lower mass accretion rates and lower photoevaporative loss rates in M dwarfs are accompanied by lower disk masses. On this topic, {\bf Barbara Ercolano} from the Institute of Astronomy, Cambridge, presented a very interesting color-color diagram to observationally distinguish disks around M dwarfs that are cleared inside-out (as expected from photoevaporation) and those that have been subject to uniform draining, due for instance to viscous evolution. Comparison of this theoretical diagram, obtained by extensive radiative transfer modelling, to available photometric observations of young stellar objects in nearby star forming regions, strongly supports a dispersal mechanism that operates from the inside out over a rapid transition timescale \citep{2010MNRAS.tmp.1505E}.

External photoevaporation induced by high-energy photons from OB stars in a cluster might be another important dispersal mechanism especially for disks around low-mass M dwarfs which have weak gravity fields. \citet{2004ApJ...611..360A} calculated that disks around M dwarfs are evaporated
and shrink to disk radii  $\le$15\,AU on short timescales $\le$10\,Myr under moderate FUV fields which are present in small stellar groups and clusters. Disks around sun-like stars are more durable and would require a much more intense FUV field  (10 times higher) to evaporate and shrink on a similarly short ($\sim$10\,Myr) timescale.

%Thus, disk photoevaporation (from the central star as well as externally induced) should be %included in planet formation models around M-dwarfs to properly evaluate the timescale and %likelihood of forming planets.

In summary, disk evolution and dispersal around isolated M dwarfs is expected to proceed as for sun-like stars. This similarity means that at least the early stages of planet formation are also likely to proceed in a similar way (as suggested in Sect.~\ref{sect:evolution}). However, as discussed below, disk dispersal is a crucial ingredient in planet formation models and both viscous evolution and (internal and external) photoevaporation should be included to properly evaluate the timescale and likelihood of forming planets.

\subsection{Planet Formation around M Dwarfs}\label{sect:pftheory}
{\bf Greg Laughlin} from the University of Santa Cruz provided a comprehensive review of the theory of planet formation around M dwarfs. First, he pointed out the main characteristics of planets detected around M dwarfs (see also Sect.~\ref{sect:planets}): a clear paucity of close-in giant planets (thought exceptions exist, see the case of Gliese 876  by \citealt{Rivera10}), mostly low eccentricities for the detected planets, and a possible correlation between the metallicity of the M-dwarfs and the detection of giant planets around them \citep{JohnsonApps09}. This last relation extends up to the more massive sun-like stars \citep{FischerValenti05} and points to giant planet formation via core accretion as the most likely formation scenario. 

According to the core-accretion paradigm giant planet formation can be broken up into 3 phases. In the first phase, the growing planet consists mostly of solid material. The planet experiences runaway growth until the feeding zone is depleted. Solid accretion occurs much faster than gas accretion in this phase. During the second phase, both solid and gas accretion rates are small and nearly independent of time. Finally, in the last stage runaway gas accretion occurs (starting when the solid and gas masses are roughly equal). The baseline model by \citet{1996Icar..124...62P}  predicted  a long ($\sim 10$\,Myr) timescale to form a planet like Jupiter at 5\,AU. This timescale is clearly too long when compared to the observed disk lifetimes (Sect.~\ref{sect:evolution}) hence several ideas have been proposed to shorten the longest of the phases,  the core mass accretion followed by gas accretion. 

\citet{2004ApJ...612L..73L} have extended the core-accretion theory of planet formation around M dwarfs. They showed that giant planet formation is highly suppressed around  a 0.4 solar mass star in comparison to sun-like stars due to the lower disk surface densities beyond the snow line and  longer timescales both for the formation of planetesimals as well as for their accumulation. Hence, the zeroth order predictions from the core-accretion theory seem to be correct: higher metallicity -- more planets (lower metallicity -- fewer planets); higher stellar mass -- more planets (lower stellar mass -- fewer planets). The case of the planetary system around Gliese 581 (at least four planets comprising super-Earths and hot Neptunes) suggests that the planet formation process responsible for the majority of the planets in the galaxy occurs robustly in M-star systems. The formation of super-Earths has not been investigated in detail. Simulations from \citet{2007ApJ...669..606R} showed unambiguously  that if the protoplanetary disks associated with the lowest mass stars are scaled down versions of the Sun's protoplanetary disk, then the resulting terrestrial planets rarely exceed the mass of Mars, are dry, and inhospitable. 
\citet{2009Icar..202....1M} investigated a more optimistic scenario in which the disk mass is largely independent of the  stellar mass and showed that  habitable planets can form easily even around a 0.1\,M$_\odot$ star. Observations of disk masses are an essential input to the theory as discussed in Sect.~\ref{sect:properties_disks}.

%\begin{figure}[!ht]
%\plotone{MdwarfPlanetForm.ps}\label{fig:planetform}
%\caption{Accretion simulation of planet formation around a 0.12 $M_{\odot}$ late-type %M-%dwarf (Montgomery \& Laughlin 2009). The simulation ends with planets of mass 0.86, %0.79, %1.03, 0.36 and 0.22 $M_{\oplus}$}
%\end{figure}

%\begin{figure}[!ht]
%\epsscale{0.3}
%\plotone{mr-cs16.ps}\label{fig:massradius}
%\caption{Mass-radius relationship which includes the new binaries discovered as part of the %WFCAM Transit Survey.}
%\end{figure}

\section{Planets around M Dwarfs}\label{sect:planets}
We started this session with the contributed talk by {\bf Jayne Birkby} from
the Institute of Astronomy, Cambridge. She presented the first results
from the WFCAM Transit Survey carried out with the 3.8m UKIRT at
infrared wavelengths. The main goal of the survey is to place
meaningful observational constraints on the occurrence of rocky and
giant planets around low-mass stars using time-series photometric
observations of a large ($\sim6,000$) sample of M dwarfs. The survey
is half-way to completion, with extensive scrutiny and follow-up
already performed on approximately one third of the M-dwarf sample. So far, the
survey shows no hot Jupiter detections nor the more commonly expected
hot-Neptune detections. However, with only a third of the sample
thoroughly explored at this time, it is too early to speculate on the
planet fraction for the M-dwarf sample. The team
also reported the detection of a number of M-dwarf eclipsing binary
systems with masses $<0.6 M_{\sun}$ which can be used to further test
the mass-radius relation, and they have found a very low mass
($\sim0.2+0.1M_{\sun}$) eclipsing binary, which is near the theorized
limiting mass for formation via disk fragmentation.

%We started this session with the contributed talk by {\bf Jane Birky} from the Institute of %Astronomy, Cambridge. She presented the first results from the WFCAM Transit Survey carried %on with the 3.8m UKIRT at infrared wavelengths. The main goal of the survey is to place %meaningful statistical constraints on the occurrence of planets (rocky and giants) around low %mass stars using time-series photometric observations of a large ($\sim$6,000) sample of %M-%dwarfs. At just over the half-way stage of the survey, they reported the detection of a %
%number of M-dwarf eclipsing binary systems with masses $<$0.6\,M$_\odot$ which can be %used to further test the mass-radius relation. In addition, they found a very low mass %($\sim$0.2$+$0.1\,M$_\odot$) eclipsing binary. So far, they did not detect Jupiters around %M-%dwarfs, maybe due to the paucity of these planets around very low-mass stars, but have the %sensitivity to detect Neptune-size planets, which should be common around M-dwarfs. 

We concluded our CS16 splinter session with the review talk by {\bf Scott
Gaudi} from The Ohio State University on the demographics of planets
around M dwarfs. Although the faintness of M-dwarf stars makes it hard to
detect planets around them with almost any technique, their smaller
mass and size present some advantages. In the case of the radial
velocity (RV) technique, a giant planet around a M dwarf will produce
a larger radial velocity signal than the same planet around a sun-like
star. However, precision RV surveys in the optical are limited to
bright early M dwarfs (e.g., \citealt{2009A&A...507..487M}, \citealt{2010PASP..122..905J},
\citealt{2010ApJ...723..954V}).  Infrared RV searches are just starting and are
promising especially for detecting planets around the very low-mass
M dwarfs (e.g., \citealt{2010ApJ...713..410B}) . In the case of transits, depths of
transits due to
planets orbiting M dwarfs are larger ($\sim 10\%$ for Jupiter-size
planets), but the duty cycles and transit probabilities are lower for
fixed period. On the other hand, for habitable planets, the transit
depths, probabilities, and duty cycles are dramatically larger for
low-mass stars, thus favoring the discovery of habitable planets
around M dwarfs \citep{2003ApJ...594..533G}.  The main challenge with
discovering transiting planets around M dwarfs is their intrinsically
low-luminosity.  Surveys for transiting planets around bright M dwarfs
such as MEarth \citep{2008PASP..120..317N,2009Natur.462..891C}
must contend with the paucity and sparsity of targets (i.e., there are
only 2,000 M-dwarfs with $V\la 12$ over the entire sky), whereas
deep, pencil-beam surveys (see the contribution from J. Kirby) must
contend with the difficulties with RV follow-up for the faint
candidates.  Microlensing has the advantage of being insensitive to host
star luminosity and thus sensitive to planets throughout the Galaxy.
Furthermore, since M dwarfs are the most common stars in the Galaxy,
the microlensing host stars are likely to be M-dwarf stars. Although the
faintness of the host star presents challenges for characterization of
the detected systems, often the host star and planet masses can also
be measured with followup observations to within $\sim 10-20\%$
\citep{2007ApJ...660..781B}.  Microlensing is most sensitive to planets with projected
separations near the Einstein ring of the host star, which is $\sim
3.5({\rm M/M}_\odot)^{1/2}$ and thus a factor of $\sim 3$ times the snow
line for typical parameters \citep{2010ApJ...720.1073G}.  Thus microlensing
discoveries are complementary to those of close-in planets detected
via RV or transit searches.

\begin{figure}[!ht]
\plottwo{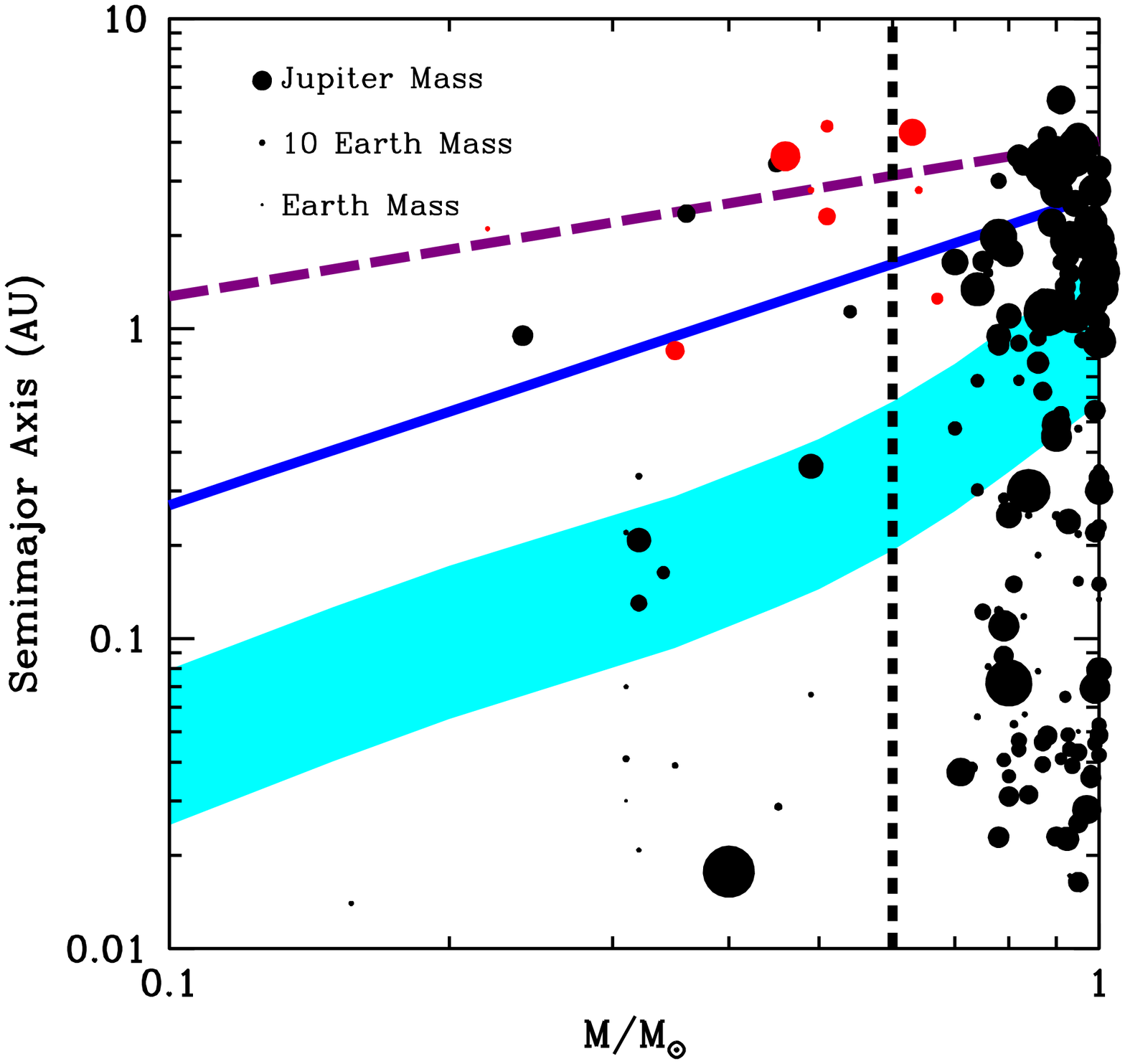}{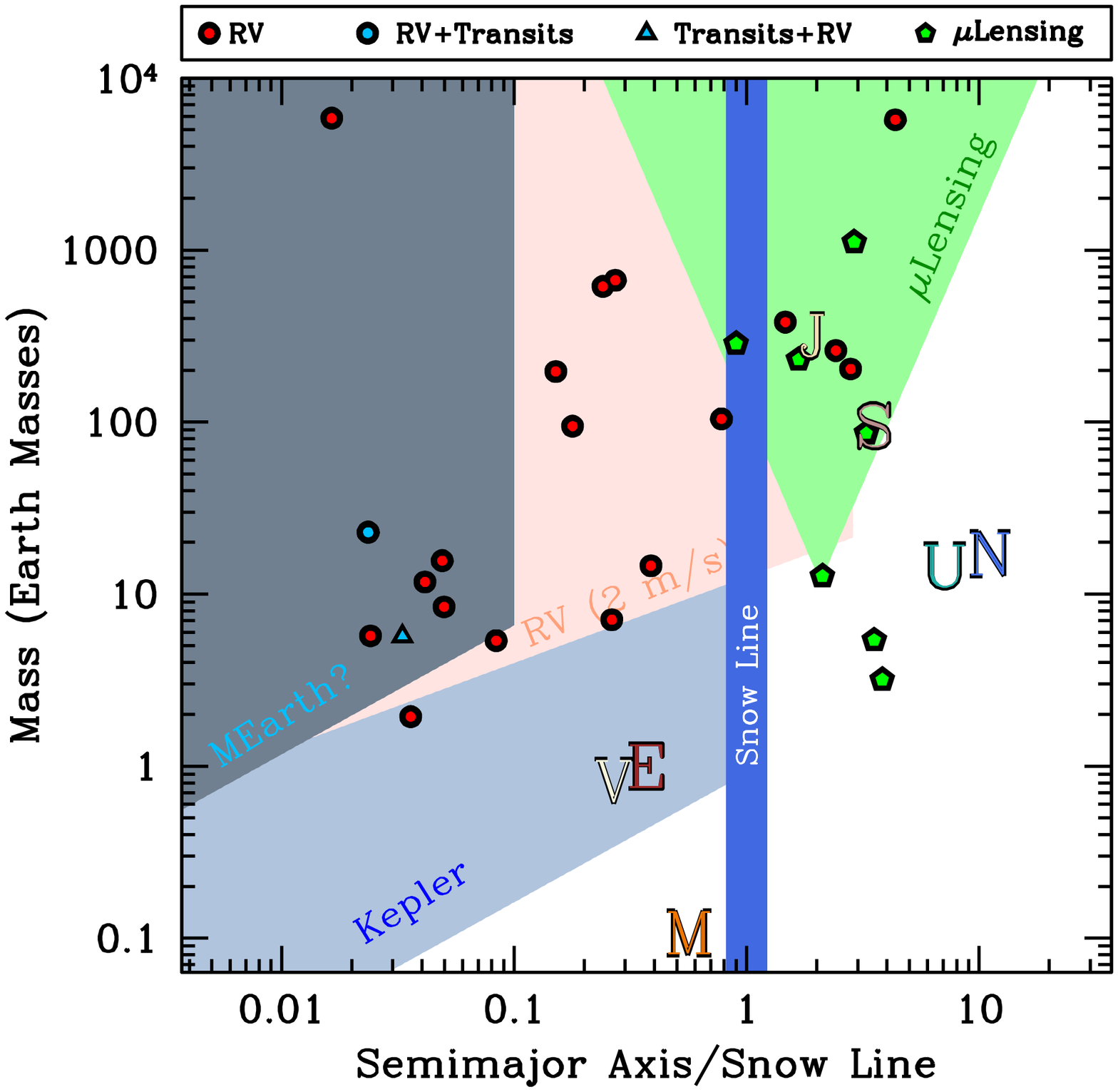}\label{fig:demo}
\caption{Left Panel: Semimajor axis versus host star mass for planets discovered by radial
velocity or transits (black), and planets discovered by microlensing
(red).  The size of the points are proportional to the mass of the
planets to the 1/3 power.  The vertical dashed line shows the
approximate mass limit for M dwarfs.  The cyan shaded region shows the
"conservative" habitable zone according to Selsis et al. (2007).  The
blue solid line is the location of the snow line, assuming a scaling
of $a_{\rm SL} = 2.7~{\rm AU}(M/M_\odot)$.  The dashed purple line is
the typical Einstein ring radius for microlens host stars. Right Panel: Planet mass versus semimajor axis in units of the snow line for the 27
planets orbiting stars with $M<0.6~M_\odot$ known at the time of the
talk. Also shown are the location of the solar system planets, as well
as approximate regions of sensitivity for various transit, radial
velocity, and microlensing surveys.}
\end{figure}

At the time of the talk 27 planets were known orbiting M-dwarf stars: 19 of
them were discovered via the RV technique, 1 via transit, and 7 via
microlensing. One system identified via microlensing is particularly
interesting since it appears to be a scaled-down version of the Solar
System \citep{2008Sci...319..927G,2010ApJ...713..837B}. Several of the
planets discovered so far present extreme planet/star mass ratios,
examples are HD41004b \citep{2003A&A...404..775Z} and the GJ876 system
\citep{2001ApJ...556..296M}, which may be challenging to explain via the
core-accretion theory \citep{2004ApJ...612L..73L}. RV searches point to a
paucity of close-in (distances $<$ a few AU) giant planets around
M dwarfs \citep{2007ApJ...670..833J} and to a possible correlation between
the presence of giant planets and the metallicity of the central star
\citep{2009ApJ...699..933J,2010A&A...519A.105S}. 
On the contrary, microlensing shows that giant planets are
relatively common around M dwarfs at radial distances of a few times
the snow line  \citep{2010ApJ...720.1073G}. These results together might
indicate that giant planets form in M-dwarf disks but do not migrate
inward. Additional possible trends discussed in the literature include
the more likely presence of low-mass planets around low-mass stars
\citep{2007A&A...474..293B} and a higher fraction of low-mass planets in
multiple systems \citep{2009A&A...507..487M}.

\section{Summary}
Because of their proximity and numbers M dwarf-stars are attractive targets to search for Earth-like planets with near-future instruments. Surveys to detect giant- and down to Super-Earth-size planets are well under way and have already identified interesting planetary systems around M dwarfs. While some of them seem to be scaled down versions of our Solar System, others are very different in planet/star mass ratio and challenge our current understanding of how planets form from the circumstellar gas and dust around young M dwarfs. It is clear that a complete understanding of planet formation does require inputs both from the demographics of planets as well as from the properties and evolution of the protoplanetary disks within which planets form. 
The evolution of protoplanetary disk masses as a function of central star mass is one of most important input parameters to planet formation theories. In addition, properly accounting for the mechanisms dispersing protoplanetary disks is necessary to understand both when and in what environment planets form as well as whether they can migrate after their formation.

%\acknowledgements 

\bibliography{pascucci_i}

\end{document}